\documentclass[12pt]{article}

\usepackage{amsmath,amssymb,amsfonts,amsbsy}
\usepackage{cite}
\usepackage{graphicx}
\usepackage{wrapfig}
\usepackage{epsfig}

\usepackage{bbm} 
\usepackage{bm} 
\usepackage{color}                                                       %
\usepackage{dsfont} 
\usepackage{latexsym} 
\usepackage{lscape} 
\usepackage{mathrsfs} 
\usepackage{morefloats} 
\usepackage{floatflt} 
\usepackage{slashed} 
\usepackage{psfrag}

\DeclareFontFamily{OT1}{mygreek}{}%
\DeclareFontShape{OT1}{mygreek}{m}{n}{<->omsegr}{}%
\DeclareFontShape{OT1}{mygreek}{b}{n}{<->omsegrb}{}%
\DeclareFontShape{OT1}{mygreek}{m}{it}{<->omsegri}{}%
\DeclareFontShape{OT1}{mygreek}{bx}{n}{<->sub * mygreek/b/n}{}%
\DeclareFontShape{OT1}{mygreek}{m}{sl}{<->sub * mygreek/m/it}{}%
\DeclareSymbolFont{Greekrm}{OT1}{mygreek}{m}{n}
\DeclareSymbolFont{Greekbf}{OT1}{mygreek}{b}{n}
\DeclareSymbolFont{Greekit}{OT1}{mygreek}{m}{it}
\DeclareMathSymbol{\omegab}{\mathalpha}{Greekbf}{119}

\begin{document}
\addcontentsline{toc}{subsection}{{Transverse spin and transverse 
momentum structure of the nucleon from the COMPASS experiment}\\
{\it F. Bradamante}}
\graphicspath{{exper/yourname/}}

\setcounter{section}{0}
\setcounter{subsection}{0}
\setcounter{equation}{0}
\setcounter{figure}{0}
\setcounter{footnote}{0}
\setcounter{table}{0}

\begin{center}
\textbf{TRANSVERSE SPIN AND TRANSVERSE MOMENTUM STRUCTURE OF THE NUCLEON FROM THE COMPASS EXPERIMENT}

\vspace{5mm}

\underline{F. Bradamante}$^{\,1 \, \dag}${\\on behalf of the COMPASS 
Collaboration}

\vspace{5mm}

\begin{small}
  (1) \emph{Dipartimento di Fisica, Universit\`a degli Studi di Trieste} \\
  $\dag$ \emph{E-mail: Franco.Bradamante@cern.ch}
\end{small}
\end{center}

\vspace{0.0mm} 

\begin{abstract}
A selection is presented of recent results from the COMPASS Collaboration 
on transverse spin and transverse momentum effects in semi-inclusive 
deeply inelastic scattering (SIDIS) of 160 GeV/c muons off proton and 
deuteron targets. 
\end{abstract}

\vspace{7.2mm}

\section{Introduction}
The description of the partonic structure of the nucleon is one of the 
central problems of hadronic physics. In recent years considerable 
theoretical and experimental progress has been made and the relevance of 
the quark transverse spin and transverse momentum has been clearly 
assessed. In the present theoretical framework, eight transverse 
momentum dependent parton distribution functions (TMD PDFs) are required 
at leading twist for each quark flavour. 
They describe all possible 
correlations between the transverse momentum and spin of the quarks and 
the spin of the nucleon. 
When integrating over the quark transverse 
momentum five of these functions vanish, while three of them give the 
well known number, helicity and transversity distribution functions. 
Among these last three functions, the transversity distribution, which 
is the analogous of the helicity PDFs in the case of transversely polarized 
nucleons, was thoroughly studied only in the â€˜90s and experimentally it 
is the least known one. 
On the experimental side, 
semi-inclusive deeply inelastic lepton scattering 
(SIDIS) is  today the major source of information to access the TMD PDFs. 
It allows to access easily convolutions of the different TMD PDFs and 
fragmentation functions via high statistic measurements of asymmetries 
in the azimuthal distributions of the final-state hadrons. Also, using 
different ($p$, $d$, or $n$) targets and identifying the final state hadrons, 
one can separate the contributions of the quarks of different flavour. 
The clear non-zero spin asymmetries recently measured in SIDIS off 
transversely polarized targets by both HERMES at DESY and COMPASS at 
CERN at different beam energies, can be described quite well with the 
present formalism, and thus give much  confidence in the overall 
picture~\cite{AidalaBarone}.

COMPASS (COmmon Muon and Proton Apparatus for Structure and 
Spectroscopy) is a fixed-target experiment at the CERN SPS taking data 
since 2002. 
The COMPASS spectrometer is by now very well known in the
scientific community and I will not spend any time in describing it,
but only refer to the  NIM paper of Ref.~\cite{Abbon:2007pq} and to the
previous speaker~\cite{Bressan}.
An important part of the experimental programme consists in 
the study of the nucleon structure and SIDIS data have been collected using a 
160 GeV longitudinally polarized muon beam and either longitudinally or 
transversely polarized proton (NH$_3$) and deuteron ($^6$LiD) targets. 
A selection of the 
results on the azimuthal asymmetries 
in $\mu N \rightarrow \mu ' h^{\pm} X$
extracted from the data collected 
with transversely polarized targets is presented, with particular 
focus on the most recent measurements from the data collected in 2007 
and 2010 with the proton target. 
These results exhibit clear signals for 
the Collins asymmetry, interpreted as a convolution of a non-zero 
transversity PDF and the Collins fragmentation function (FF), 
and for the Sivers 
asymmetry which is related to the Sivers function, the most famous and 
discussed of the TMD PDFs.  
At the same time six more transverse spin dependent azimuthal  
asymmetries have been obtained from the proton and the deuteron data. 
They have all their own interpretation in terms of the QCD parton model, 
preliminary results have already been presented at several conferences, 
but I have not enough space to include them in this written report.  
Large asymmetries have been measured in the production of 
oppositely charged hadron
pairs (2-h) and the comparison between the Collins asymmetry and the
2-h asymmetry has led to interesting observations on the hadronisation
mechanism of transversely polarised quarks.
The data collected with the $^6$LiD target, suitably averaged up to 
cancel possible target polarization effects, have also been analysed to 
search for the azimuthal modulations in the production of hadrons which 
are expected  to be present in the unpolarised SIDIS cross-section. 
The 
azimuthal hadron asymmetries, which are related to the Boer-Mulders TMD 
PDF, show strong and somewhat puzzling kinematical dependences. 

\section{Collins and Sivers asymmetries}

SIDIS data with a 160 GeV $\mu ^+$ beam
and with the transversely polarised deuteron target  ($^6$LiD) were
taken in the years 2002 to 2004.
In 2007 and 2010 the transversely polarised proton  target  (NH$_3$)
was used, again with the 160 GeV $\mu ^+$ beam.

The data analysis is very similar for all the years of data taking and 
the relevant cuts applied to select the 
``good events''  are also the same.
Only events with photon virtuality 
$Q^2 > 1$ (GeV/c)$^2$, fractional energy of the virtual photon
$ 0.1 < y < 0.9$, and mass of the hadronic final state system 
$W > 5$ GeV/c$^2$ are considered.
The charged hadrons are required to have at least 0.1 GeV/c transverse 
momentum $p_T^h$
with respect to the virtual photon direction and a fraction 
of the available energy $z > 0.2$. 
\begin{figure}[tb]
\begin{center}
\includegraphics[width=0.49\textwidth]{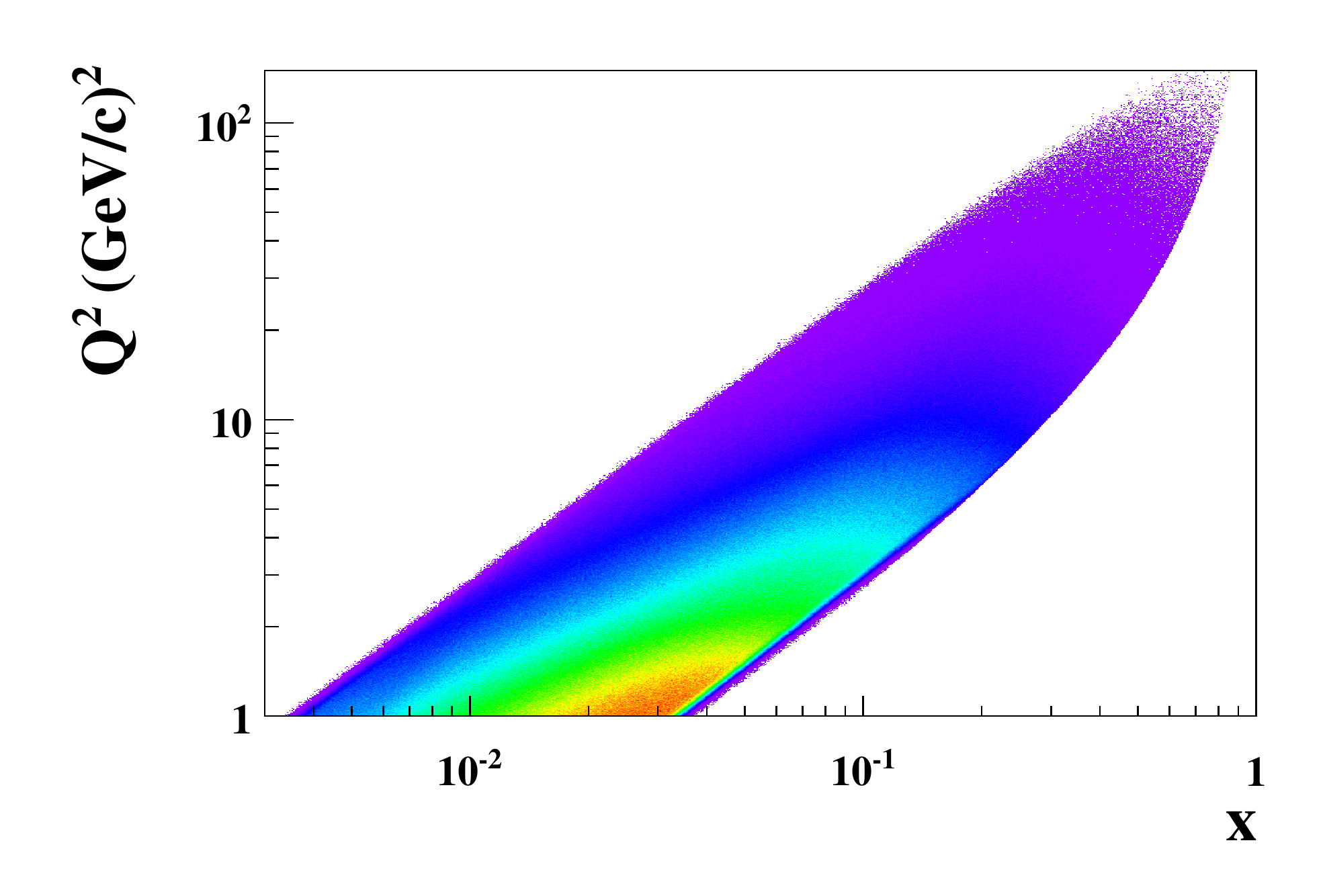}
\includegraphics[width=0.49\textwidth]{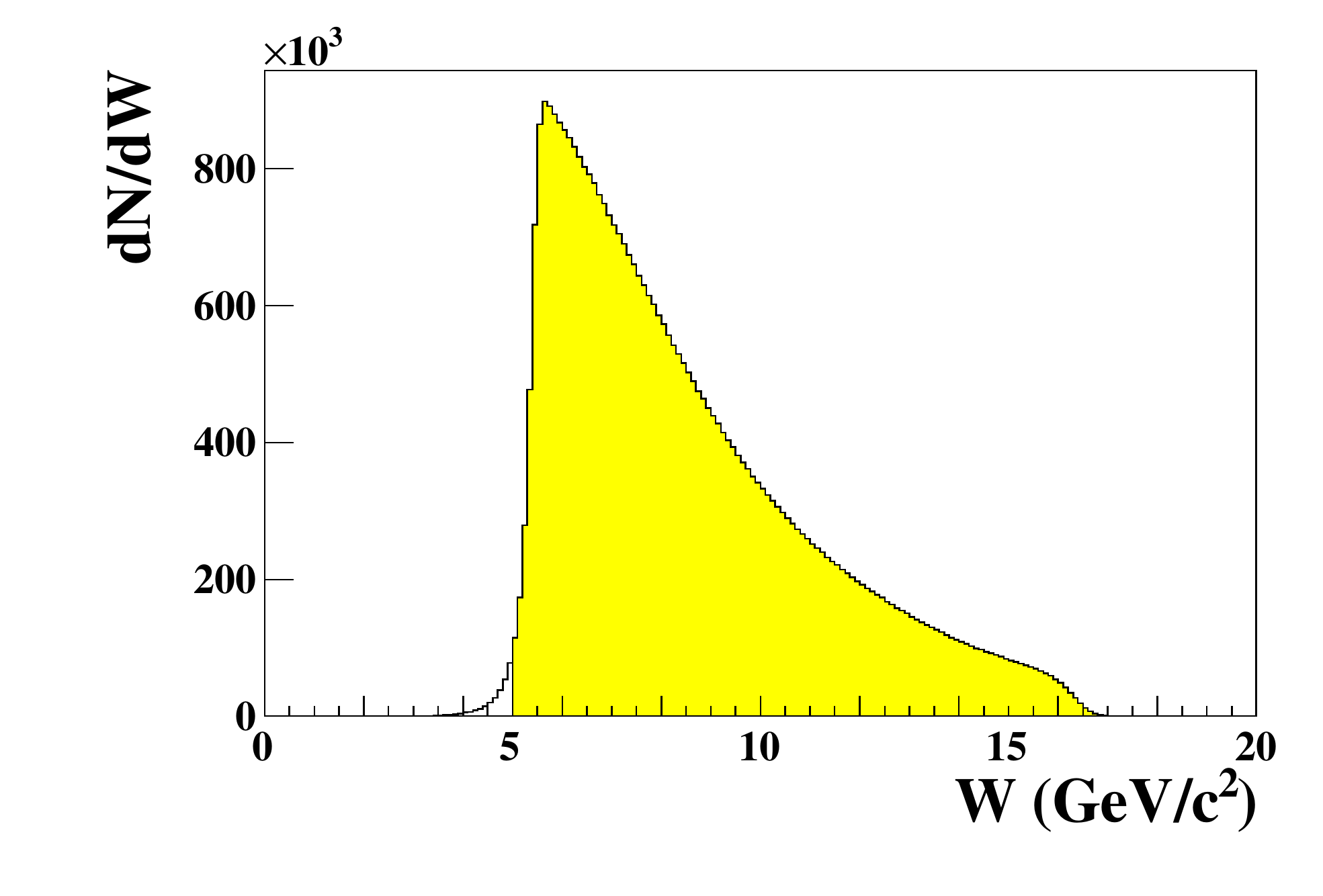}
\vspace*{-0.5cm}
 \caption{Left: $x-Q^2$ correlation for charged hadrons.
Right: $W$ distribution.
\label{fig:kin}}
\end{center}
  \end{figure}
The $x-Q^2$ correlation for charged hadrons from 2010 data
is shown in Fig.~\ref{fig:kin} 
(left).
As can be seen, the $x$ range goes from
 $x\simeq 3 \cdot 10^{-3}$ to $x\simeq 0.7$ with relatively large
$Q^2$ values in the valence region.
Figure~\ref{fig:kin} 
(right) gives the corresponding 
$W$ distribution.
In the standard analysis, the transverse-spin asymmetries are  
measured separately for positive and negative hadrons (or pions or kaons) 
as functions of $x, \, z$ or $p_T^h$.
The complete definition (namely sign and kinematic factors)
of the asymmetries can be found in the published 
papers~\cite{Adolph:2012snsp}.

The Collins and Sivers asymmetries for positive and negative 
hadrons from the 2004 deuteron data~\cite{Alexakhin:2005iwda} 
turned out to be compatible with zero
within the few percent uncertainties, at variance
with the non-zero results obtained by the HERMES experiment on 
proton~\cite{Airapetian:2010dsae}.
These data could be understood in terms of  cancellation between 
the $u$ and $d$ quark
contributions in the deuteron target, and
together with the Belle data of the $e+e- \rightarrow hadrons$
process were used in global fits
to extract the transversity and Sivers functions.
Still today these COMPASS data are the only SIDIS data collected with
a transversely polarised deuteron target.

The first results for the charged hadrons Collins and Sivers asymmetries
on proton from COMPASS~\cite{Alekseev:2010rw} came from the analysis
of the 2007 data, while higher precision results have been obtained
from the 2010 data~\cite{Adolph:2012snsp}.
Very recently,  results for charged pions and kaons
have also been produced~\cite{Martin:2013eja}.

\begin{figure}[tb]
\begin{center}
\includegraphics[width=0.9\textwidth]{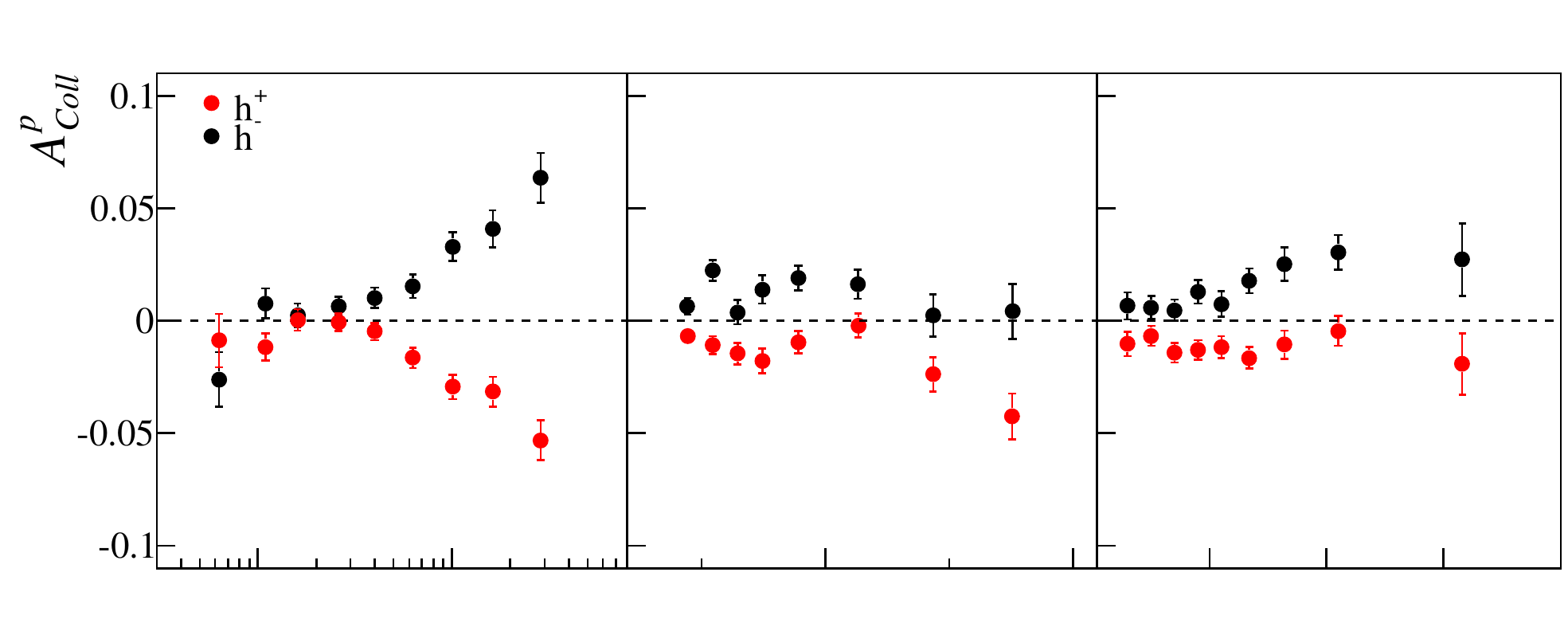}
\includegraphics[width=0.9\textwidth]{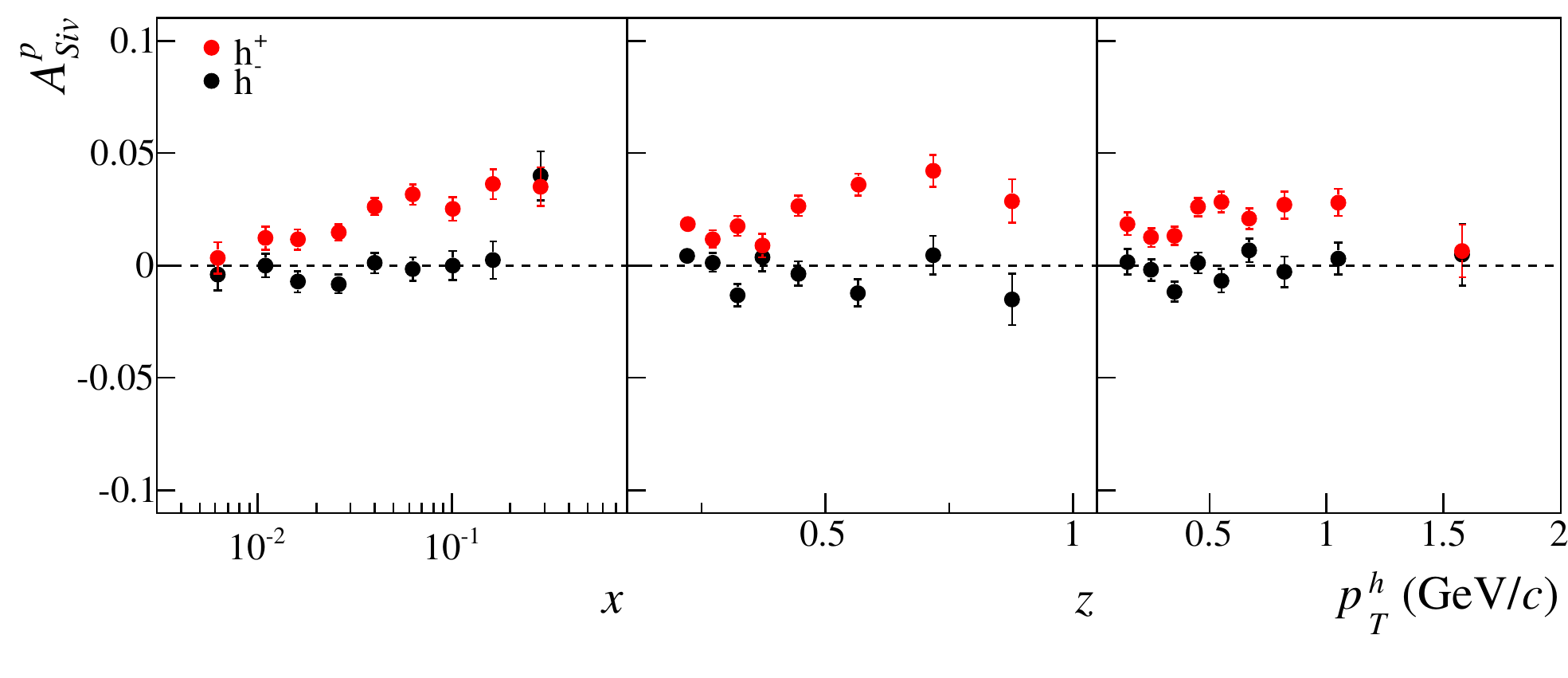} 
\vspace*{-.5cm}
 \caption{Collins (upper panel) and Sivers (lower panel)
asymmetries for positive (red points) and negative (black points)
hadrons as functions of $x$, $z$ and $p_T^h$
from the combined 2007 and 2010
 proton data.
\label{fig:colsiv}}
\end{center}
  \end{figure}
The combined results for non-identified hadrons from 2007 and 2010
are shown in Fig.~\ref{fig:colsiv}.
The Collins  asymmetries (upper plots)
are compatible with zero in the previously unmeasured $x<0.03$ region while 
at larger $x$ they are clearly different from zero, 
with opposite sign for positive and 
negative hadrons and in nice agreement, both in sign and in
magnitude, with the HERMES results~\cite{Airapetian:2010dsae}.
There is no indication for lower values of the Collins 
asymmetry at the higher COMPASS
$Q^2$ values as compared to the HERMES measurement.

The Sivers asymmetries for charged hadrons are 
given in the lower plots of Fig.~\ref{fig:colsiv}.
For $h^-$ they are compatible 
with zero with some indication for small negative values over the entire
$x$ range
but in the last bin.
In the case of $h^+$, the Sivers asymmetry is positive down 
to very small $x$  values and  in the $x>0.03$ region
it is smaller than the same asymmetry
measured by HERMES~\cite{Airapetian:2010dsae},
a fact which can be understood in terms of the recent calculations
on TMDs evolution.

\section{Two-hadron asymmetry}
\label{sec:2ha}
An alternative approach to the transversity PDF in SIDIS utilises 
the transverse spin asymmetry in the production of pairs of oppositely 
charged hadrons, in the process  
$lN \rightarrow l’ h^+h^- X$~\cite{Collins:1993kq}.  
In the 
SIDIS cross-section an azimuthal modulation is expected as a function of 
$\phi_{RS}= \phi_R + \phi_S - \pi$, whose amplitude is proportional to 
the product 
of the transversity PDF and a new chiral-odd FF, the Dihadron Fragmentation 
Function (DiFF) $H^{\angle}(z,M_{h+h-},\cos \theta)$~\cite{Bacchetta:2002ux}. 
\begin{figure}[tb]
\begin{center}
\includegraphics[width=0.98\textwidth]{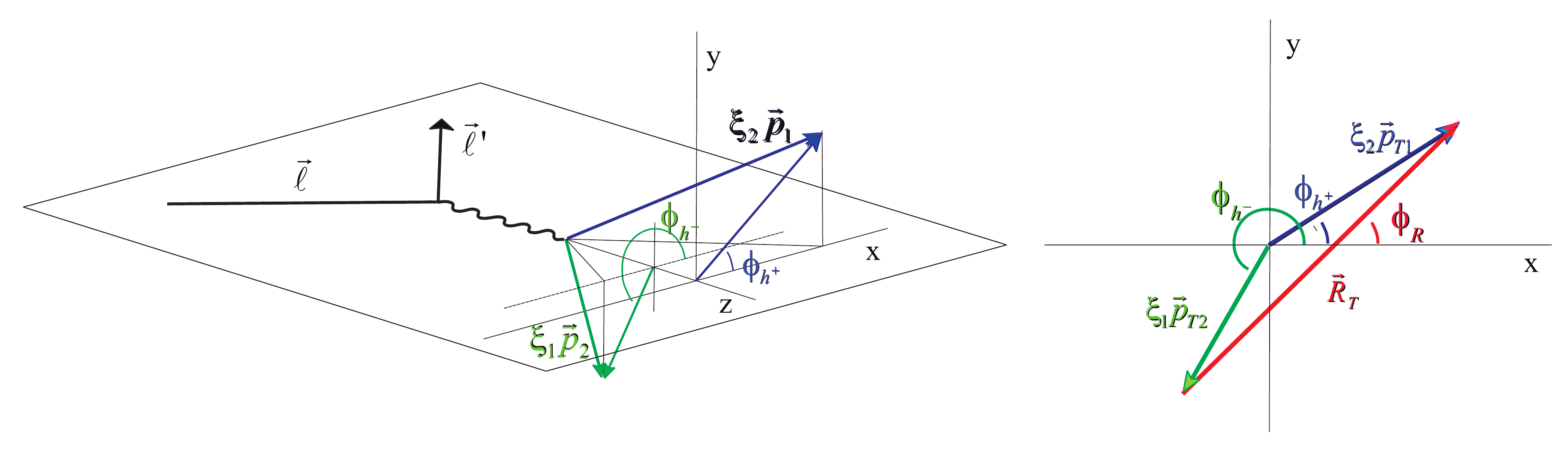}
 \caption{Kinematics of hadron pair production  process in SIDIS.
The 3-momenta $\vec{\ell}$ and $\vec{\ell}'$ of the incoming and scattered lepton
define the scattering plane, the z axis (the direction of
the virtual photon direction) and the x axis.
The vectors $\vec{p}_1$ and $\vec{p}_2$ are the 3-momenta of the
positive and negative hadron respectively.
The vector  $\vec{R}$ is defined as 
$\vec{R}=(z_2 \, \vec{p}_1-z_1 \, \vec{p}_2)/(z_1+z_2)=
\xi_2 \, \vec{p}_1-\xi_1 \, \vec{p}_2$. The subscript $T$ indicates
the transverse component with respect to the virtual photon direction.
\label{fig:2hkin}}
\end{center}
  \end{figure}
The angle $\phi_R$ is the azimuthal angle of the relative momentum $R$ of 
the two hadrons  as depicted in Fig.~\ref{fig:2hkin}, 
$\pi - \phi_S$ is the azimuthal angle of 
the spin vector of the struck quark,
$z$ is the sum of the fractional 
energy  of the two hadrons, $M_{h+h-}$ is the invariant mass of the two 
hadrons,
and $\theta$ is the polar angle of $h^+$.
First evidence for azimuthal asymmetries in leptoproduction of $\pi^+\pi^-$  
pairs on transversely polarized protons was published by 
HERMES~\cite{Airapetian:2008sk}, 
while results on both proton and deuteron targets for unidentified charged 
hadrons pairs $h^+h^-$  have been published by COMPASS~\cite{Adolph:2012nw}.
Using these data and the Belle data on $e^+e^-$ annihilation into two pairs 
of hadrons~\cite{Vossen:2011fk} a first extraction of the $u$ and $d$ quark 
transversity could be 
performed~\cite{Bacchetta:2012ty}, which was in good agreement 
with the extraction of 
Ref.~\cite{Anselmino:2013vqa}, 
based on the Collins asymmetry of single hadrons.  
The same procedure was applied to directly extract  $u$ and $d$ quark 
transversities in the different x bins using COMPASS proton and deuteron 
results~\cite{CElia}.
The data collected by COMPASS in 2010 on the transversely polarized 
proton target provided a sample of hadron pairs larger than the published 
one by a factor of three. 
Preliminary results were first shown at Transversity 
2011~\cite{Braun:2012zzb}. 
The selection of the two hadron events follows the same track than the 
single-hadron analysis, but more requirements are imposed. 
All possible combinations of oppositely charged hadron pairs originating 
from the vertex are taken into account in the analysis.  
At least three outgoing tracks are demanded for an interaction vertex, 
and each hadron has to have a fractional energy $z>0.1$ and $x_F>0.1$, 
to ensure 
that the hadrons are not produced in the target fragmentation. 
A cut of $R_T>0.07$ GeV/c  ensures a good definition of $\phi_{RS}$. 
Within measurement errors, the 2-h asymmetry of the COMPASS deuteron 
data from the 2002-2004 runs are compatible with zero. 
\begin{figure}[tb]
\begin{center}
\includegraphics[trim=0cm 0.0cm 0cm 0.05cm, clip, width=0.95\textwidth]{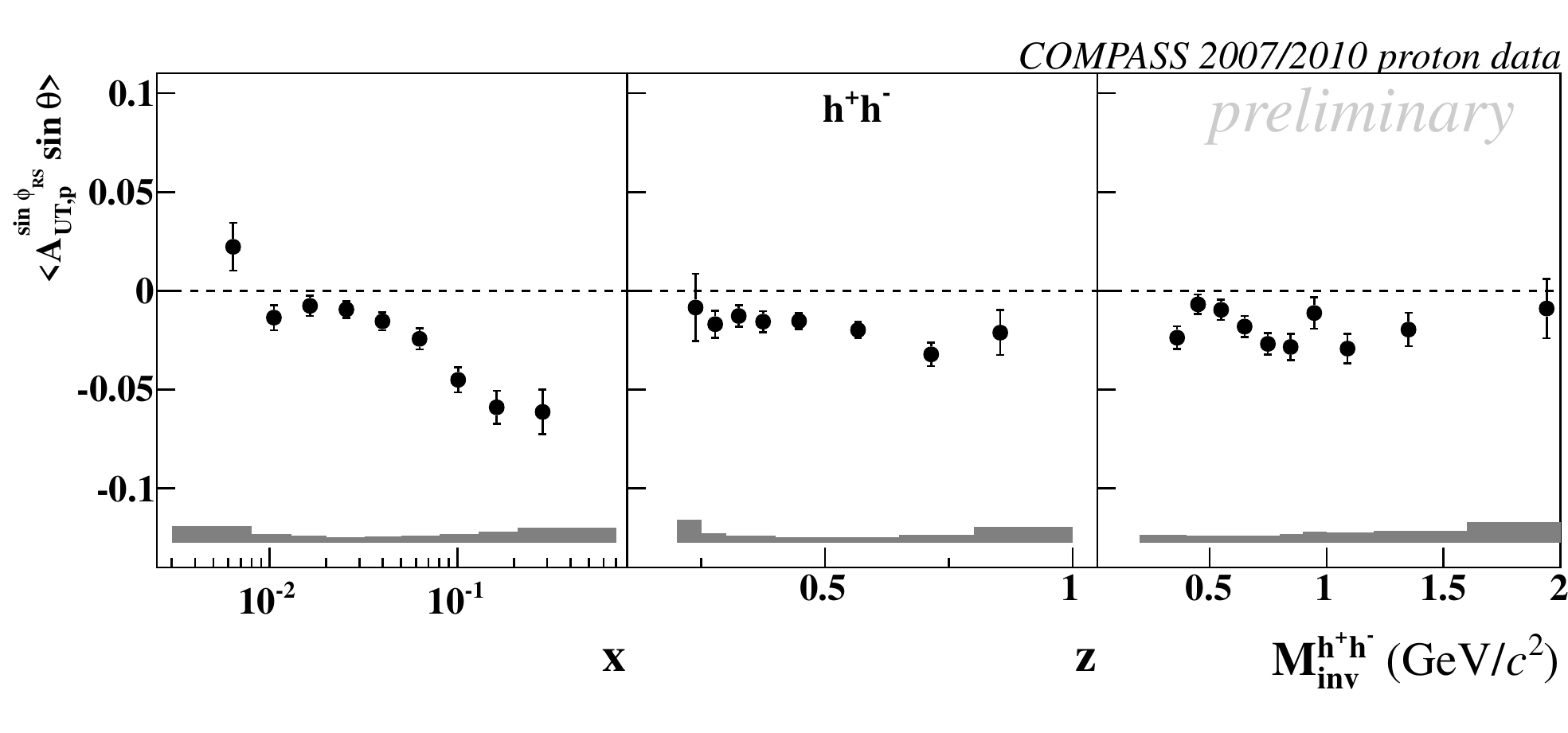}
\vspace*{-0.5cm}
 \caption{Unidentified 2-h asymmetries as functions of
$x$, $z$ and $M_{h+h-}$ from the  2007 and the 2010
 proton data.
\label{fig:2hasymm}}
\end{center}
  \end{figure}
On the other hand, 
the data on the proton target definitively show a non-zero signal, 
in particular in the $x$-valence region, as clear from Fig.~\ref{fig:2hasymm}, 
where the combined  results from the 2007 and the 2010 runs are shown
as a function of $x$, $z$ and $M_{h+h-}$. 
These data are in good agreement with the only other existing measurement 
from the HERMES Collaboration, but the statistics of the COMPASS sample 
is considerably higher thanks to the larger phase space available. 
A remarkable similarity can thus be noted between the 2-h asymmetry and 
the Collins asymmetry for $h^+$, which will be further discussed 
in  Section~\ref{sec:interp}.

\section{Azimuthal modulations in unpolarised SIDIS}

Since the early times of the quark-parton model it was realised that
a possible intrinsic transverse momentum of the target quark would cause
 measurable effects in the SIDIS cross-section, namely 
a $\cos \phi_h$ and a $\cos 2\phi_h$ modulation.  
Recently the study of these modulations has become particularly 
interesting within the framework of the new TMD approach to the PDFs and FFs. 
The amplitudes of these modulations,  
$A^{UU}_{\cos\phi_h}$ and $A^{UU}_{\cos2\phi_h}$ are not only due to the 
kinematic of the scattering process (Cahn effect) but depend also on a 
new TMD PDF, the so-called Boer-Mulders function, which describes 
the correlation between the quark transverse spin and its transverse 
momentum in an unpolarised nucleon. 
In the amplitudes the Boer-Mulders function is convoluted with the 
Collins function, and its extraction from the unpolarised SIDIS 
cross section data is an important goal of the HERMES, CLAS and 
COMPASS Collaborations.
COMPASS has extracted~\cite{gsspin12} the amplitudes $A^{UU}_{\cos\phi_h}$ and 
$A^{UU}_{\cos2\phi_h}$ from a sample of data collected in 2004 on a 
$^6$LiD target (to a good approximation, an isoscalar deuteron 
target). 
An  $A^{LU}_{\sin\phi_h}$ asymmetry is also expected to be present due to 
higher twist effects and has been measured. It has no clear interpretation 
in terms of the parton model, it turns out to be small, and will be 
neglected in the following. 
To extract the azimuthal asymmetries one has to correct the measured 
azimuthal distributions by the $\phi_h$ dependent part of the apparatus 
acceptance  and to fit the corrected distribution with the appropriate 
$\phi_h$ function. To reduce as much as possible the acceptance corrections, 
some tighter cuts have been applied to the SIDIS event selection as 
compared to the standard analysis. The final event and hadron selection 
is in this case: $Q^2 > 1$~GeV$^2$/$c^2$, \,\, $W > 5$~GeV/$c^2$, \,\, $0.003 < x < 0.13$, \,\, $0.2 < y < 0.9$, \,\,  ,
$\theta_{\gamma ^*} ^{lab} < 60$~mrad, $0.2 < z < 0.85$ and \,\, $0.1 <p_T ^h < 1.0$~GeV/$c$. 

\begin{figure*}[tbh!]
\begin{center}
\includegraphics[width=0.95\textwidth]{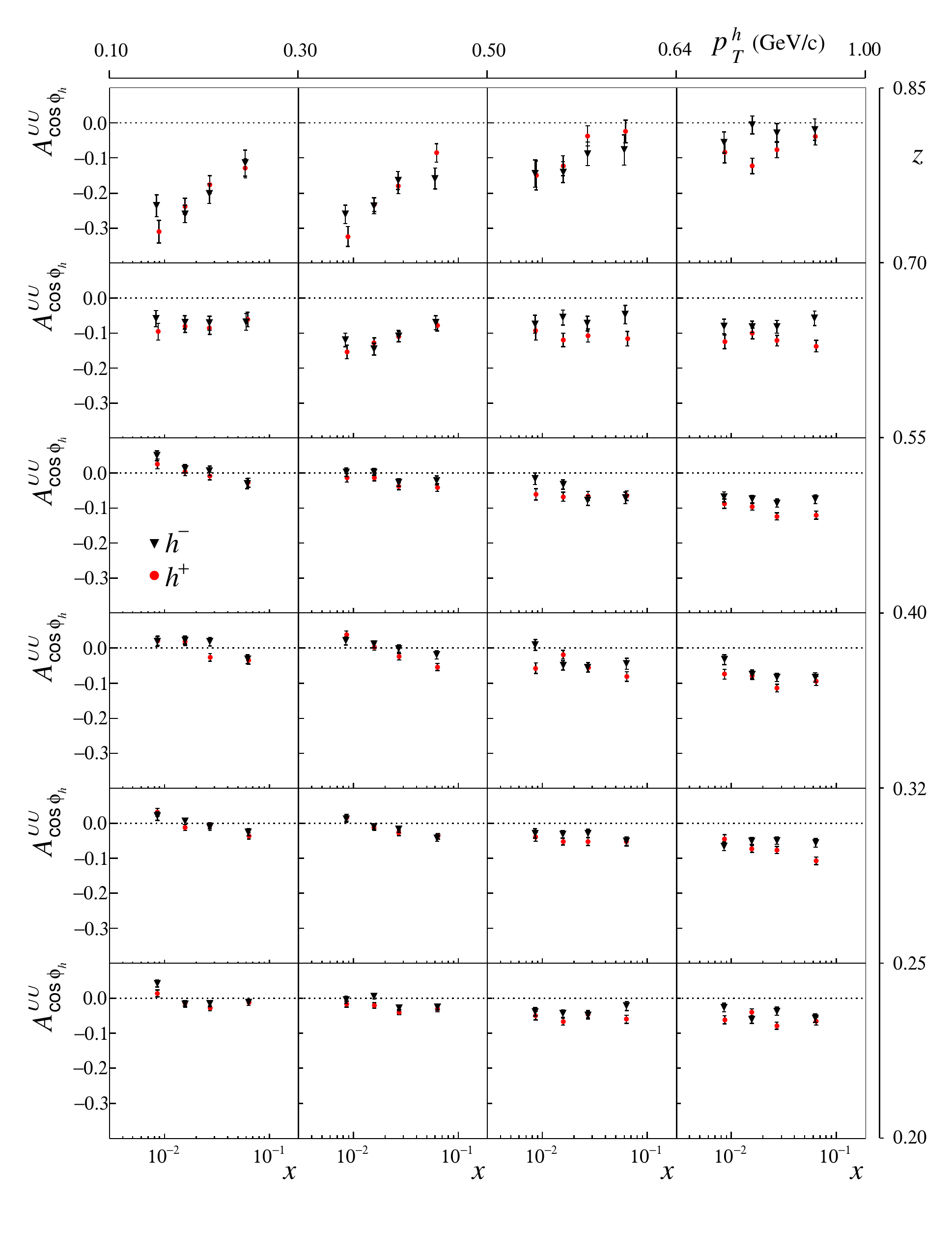}
\vspace*{-1.5cm}
 \caption{The preliminary results for the
$\cos \phi_h$ asymmetries $A^{UU}_{\cos \phi_h}$ for positive 
and negative hadrons as functions of $x$ in the different 
$z$ and $p_T^{\,h}$ bins from 2004 unpolarised $^6$LiD data.
\label{fig:cosasymm}}
\end{center}
  \end{figure*}

The amplitudes of the azimuthal modulations  have been obtained 
binning the data both separately in each of the relevant kinematic 
variables $x$, $z$ or $p_T^{\,h}$  and  in a three-dimensional grid of 
these three variables.  
The amplitudes of the $\cos \phi_h$ and  $\cos 2\phi_h$ modulations show 
strong kinematic dependences both for positive and negative hadrons. 
As an example, the preliminary results for $\cos \phi_h$ are shown in 
Fig.~\ref{fig:cosasymm}. 
Also $A^{UU}_{\cos 2\phi_h}$ shows a similarly strong dependence on the 
$x$, $z$ and $p_T^{\,h}$ variables, which up to now has not been reproduced 
with theoretical models.

\section{Interplay between Collins and dihadron asymmetries}
\label{sec:interp}
There is a striking similarity between the Collins asymmetries 
in Fig.~\ref{fig:colsiv} and the 
2-h asymmetries as functions of $x$ shown in 
Fig.~\ref{fig:2hasymm}.
First of all there is a mirror-symmetry between the Collins asymmetry for 
positive and for negative hadrons, the magnitude of the asymmetries being 
essentially identical and the sign being opposite in each x-bin. 
This symmetry has been phenomenologically described in terms of $u$ 
quark dominance 
and almost opposite 
favoured and unflavoured Collins FFs~\cite{Anselmino:2013vqa}.

The observation that the new COMPASS results on the 2-h asymmetries
allow is that the values of 
the 2-h asymmetries are slightly higher but very close to the values 
of the Collins asymmetries for positive hadrons and to the mean of the 
values of the Collins asymmetry for positive and for negative hadrons, 
after changing the sign of the asymmetry of the negative hadrons. 
The hadron samples on which these asymmetries are evaluated are different, 
since at least one hadron with $z>0.2$ is required to evaluate the Collins 
asymmetry while all the combinations of positive and negative hadrons with 
$z>0.1$ are used in the case of the 2-h asymmetry. 
It has been checked however that the similarity between the two different 
asymmetries stays the same when measuring the asymmetries on the common 
hadron sample. 
This gives a strong indication that the analysing powers of the 1- and 
2-h channels are almost the same, and their comparison
will allow to access the contribution of 
the convolution over the transverse momenta in the Collins asymmetry.

More work has been done to understand the similarities between the Collins 
and the 2-h asymmetries~\cite{fbcomo}. 
The mirror symmetry of the Collins asymmetry 
for positive and negative hadrons suggests that when a transversely 
polarized quark fragments oppositely charged hadrons have  
azimuthal angles $\phi_{h+}$ and $\phi_{h-}$ differing by $\pi$.
An anti-correlation between $\phi_{h+}$ and $\phi_{h-}$ is expected as
a consequence of the local transverse momentum conservation in the 
fragmentation. 
The new and relevant point  is that this correlation shows up also in the 
Collins asymmetry, 
so that the asymmetry exhibited by the hadron pair can be obtained
in a way which is different from the one described in Sect.~\ref{sec:2ha}.
For each pair of oppositely charged hadrons, 
using the unit vectors of their transverse momenta we have evaluated 
the angle $\phi_{2h}$ which is the arithmetic mean 
(modulus $\pi$) of the azimuthal angles of 
the two hadrons
after correcting $\phi_{h-}$ for the already mentioned $\pi$ difference.

\begin{figure}[tb]
\begin{center}
\includegraphics[width=0.95\textwidth]{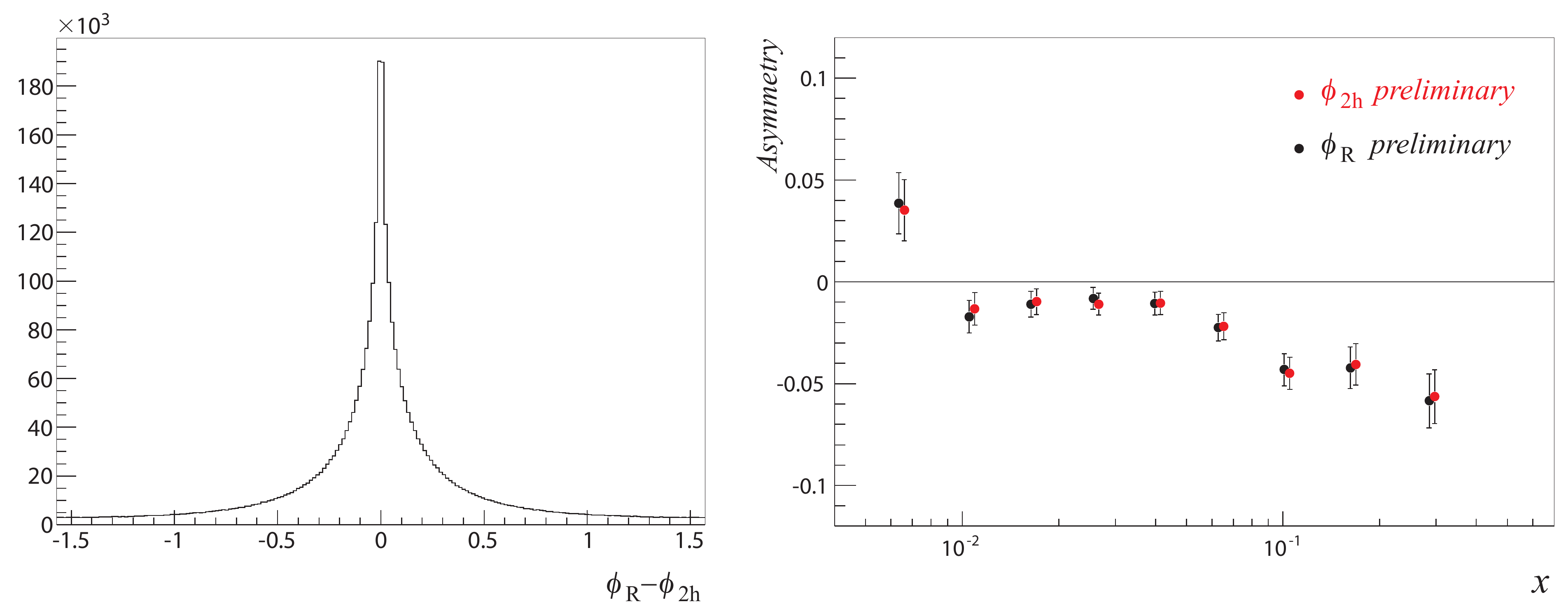}
 \caption{Left: difference between $\phi_{2h}$ and $\phi_R$.
Right: comparison between the 2-h asymmetries evaluated (see text) using
$\phi_R$ (black points) and $\phi_{2h}$ (red points) from the 2010 data.
\label{fig:fig78}}
\end{center}
  \end{figure}

This azimuthal angle of the hadron pair is strongly correlated with $\phi_R$,
as can be seen in Fig.~\ref{fig:fig78} (left) where the difference of 
the two angles is shown. 
By subtracting from $\phi_{2h}$ the azimuthal angle
$\pi - \phi_S$ as done in the standard
analysis described in Sect.~\ref{sec:2ha}, one  obtains the 
angle $\phi_{2hS}$ which is simply the
mean of the Collins angle of the positive and negative hadrons, namely a 
 Collins angle for the hadron pair. 
The amplitude of the  $\sin \phi_{2hS}$ modulation, 
which can be called the Collins asymmetries for the hadron pair, 
is shown as a function of $x$ in Fig. \ref{fig:fig78} (right) for all the 
$h^+ h^−$ pairs with $z > 0.1$ in the 2010 data, and compared with the 
2-h asymmetries extracted from the same data sample. 
It is clear that the asymmetries are very close, hinting at a common 
physical origin for the Collins mechanism and the dihadron fragmentation, 
as originally suggested in the original $^3$P$_0$ Lund model and in 
the recursive string fragmentation model~\cite{Artru:2002pua}.

\section{Acknowledgements}
I'm grateful to Anatoly for his kind invitation, and to the COMPASS 
colleagues, in particular C. Braun, A. Martin and G. Sbrizzai,  
for valuable discussions and contributions to  the work described
in Sect.~\ref{sec:interp}.

\end{document}